\documentclass{moriond}

\def\vep{\varepsilon}

\def\be{\begin{equation}}
\def\ee{\end{equation}}
\def\bea{\begin{eqnarray}}
\def\eea{\end{eqnarray}}

\usepackage{amsmath}

\newcommand{\Photo}{}

\begin{document}
\vspace*{4cm}
\title{Cosmic shear with Einstein rings}

\author{Pierre Fleury}

\address{Laboratoire Univers et Particules de Montpellier (LUPM), 
CNRS \& Université de Montpellier (UMR-5299),
Parvis Alexander Grothendieck, F-34095 Montpellier Cedex 05, France\\
Universit\'{e} Paris-Saclay, CNRS, CEA, Institut de physique th\'{e}orique, 91191, Gif-sur-Yvette, France}

\maketitle\abstracts{%
Cosmic shear is a key probe of modern cosmology. Amongst its challenges are shape noise and intrinsic alignments, both due to our ignorance of the unlensed shape of the source galaxies. I argue here that Einstein rings may be used as standard shapes to measure the external shear along their line of sight. In the \textit{Euclid} era, this new observable is expected to be a competitive and complementary probe of the large-scale structure of the Universe.
}

\section{Introduction and motivation: cosmic shear is difficult}

Weak gravitational lensing is one of the key probes of modern cosmology. When we observe distant galaxies, the deflection of light caused by the foreground inhomogeneities changes their apparent ellipticity as $\vep = \vep_0 + \gamma$, where $\vep_0$ denotes the intrinsic ellipticity of the source and $\gamma$ is the weak-lensing shear. They are complex numbers, encoding both amplitude and orientation.

The shear $\gamma$ may be seen as a projection of the gravitational tidal field between the observer and the source. As such, it is tangentially aligned with matter over-densities in the foreground, and hence so does the apparent shape of galaxies. In practice, such a coherent alignment is extracted using the two-point correlation function of the apparent ellipticity of galaxies, whose angular power spectrum reads~\cite{Kaiser_1992}
\begin{equation}
\label{eq:shear_power_spectrum}
C_\ell^\vep
= C_\ell^\gamma
= \left( \frac{3}{2} H_0^2 \Omega_{\rm m}\right)^2
\int_0^\infty \mathrm{d}\chi \; W^2(\chi) \,
P_{\rm m}\left(\chi, \frac{\ell+1/2}{\chi}\right).
\end{equation}
In eq.~\eqref{eq:shear_power_spectrum}, $H_0$ is the Hubble constant, $\Omega_{\rm m}$ the relative density of matter today, $\chi$ the radial comoving distance, $W(\chi)$ some weight function, and $P_{\rm m}(\chi, k)$ is the matter power spectrum at $\chi$ down the light cone. In other words, measuring $C_\ell^\vep$ gives access to a projection of the matter power spectrum, which turns out to be particularly sensitive to $\Omega_{\rm m}$ and $\sigma_8$, the amplitude of linear matter fluctuations. This technique is known as \emph{cosmic shear}. 

Cosmic shear is notoriously difficult to measure. Among the main challenges that weak-lensing surveys must face, two are due to our ignorance of the intrinsic ellipticity~$\vep_0$ of the source galaxies.
First, $\gamma\sim 10^{-2}$ while $\vep_0\sim 1$, so that on an individual galaxy, weak lensing has a signal-to-noise ratio on the order of $1\,\%$. Beating this \emph{shape noise} thus requires the observation of a very large number of galaxies.
Second, the first equality in eq.~\eqref{eq:shear_power_spectrum} supposes that the $\vep_0$ of different galaxies is uncorrelated. Yet, galaxies submitted to the same local gravitational field tend to align, a source of systematics known as \emph{intrinsic alignments}.

The aforementioned two difficulties would be alleviated if we disposed of \emph{standard shapes}, i.e. objects whose intrinsic shape could be inferred independently of weak-lensing distortions. Our research programme, inspired by Birrer et al.,\cite{Birrer+2017} aims to demonstrate that strong-lensing Einstein rings could play this role and be used as complementary probes of cosmic shear.

\section{Modelling the weak lensing of strong lensing}

An Einstein ring is the image produced by strong gravitational lensing when a background galaxy that is very well aligned with a foreground galaxy. We aim to study the perturbations to such images caused by matter inhomogeneities other than the main lens, all along the line of sight, between the observer, the lens, and the source.

Consider a source which, in the absence of any light deflection, would be observed at an angular position~$\boldsymbol{\beta}$ with respect to an arbitrary origin. Suppose that the main deflector alone has a lensing potential $\psi$, and that the line-of-sight perturbations are well modelled by tidal fields on the relevant scales of the problem. Then the lens equation, which connects the image position~$\boldsymbol{\theta}$ to the source position $\boldsymbol{\beta}$, reads~\cite{Kovner_1987}
\begin{equation}
\label{eq:tidal_LOS_model}
\boldsymbol{\beta}
=
\boldsymbol{\mathcal{A}}_{\rm os}
\boldsymbol{\theta}
-
\boldsymbol{\mathcal{A}}_{\rm ds}
\boldsymbol{\nabla}\psi
(\boldsymbol{\mathcal{A}}_{\rm od}\boldsymbol{\theta}) ,
\qquad
\boldsymbol{\mathcal{A}}_{\rm ab}
=
\begin{bmatrix}
1 - \kappa_{\rm ab} - \mathrm{Re}(\gamma_{\rm ab}) &
- \mathrm{Im}(\gamma_{\rm ab}) \\
- \mathrm{Im}(\gamma_{\rm ab}) & 
1 - \kappa_{\rm ab} + \mathrm{Re}(\gamma_{\rm ab})
\end{bmatrix} .
\end{equation}
In eq.~\eqref{eq:tidal_LOS_model}, $\boldsymbol{\nabla}$ must be understood as a gradient with respect to $\boldsymbol{\theta}$, while $\boldsymbol{\mathcal{A}}_{\rm os}, \boldsymbol{\mathcal{A}}_{\rm ds}, \boldsymbol{\mathcal{A}}_{\rm od}$ are distortion matrices that encode the effect of line-of-sight perturbations. For instance, $\boldsymbol{\mathcal{A}}_{\rm os}$ represents the weak-lensing distortions that would act on the source in the absence of the main lens ($\psi=0$); similarly, $\boldsymbol{\mathcal{A}}_{\rm od}$ encode distortions due to foreground perturbations (from the observer to the deflector) and $\boldsymbol{\mathcal{A}}_{\rm ds}$ those due to background perturbations (from the deflector to the source). Each of the three $\boldsymbol{\mathcal{A}}_{\rm ab}$ is associated with a convergence~$\kappa_{\rm ab}$ and a complex shear $\gamma_{\rm ab}$.

Suppose that we try to model the image of an Einstein ring using eq.~\eqref{eq:tidal_LOS_model}. In addition to the main-lens model and the three distortion matrices, we must also model the unobserved source. As it turns out, the latter is degenerate with $\boldsymbol{\mathcal{A}}_{\rm os}, \boldsymbol{\mathcal{A}}_{\rm ds}, \boldsymbol{\mathcal{A}}_{\rm od}$, making their individual measurements impossible.

This degeneracy may be lifted as follows.\cite{Fleury+2021} Since any source-position transformation $\boldsymbol{\beta}\mapsto\boldsymbol{\beta}'$ leads to an equally valid lens equation, we may multiply eq.~\eqref{eq:tidal_LOS_model} with $\boldsymbol{\mathcal{A}}_{\rm od}\boldsymbol{\mathcal{A}}_{\rm ds}^{-1}$, which yields
\begin{equation}
\label{eq:minimal_LOS_model}
\boldsymbol{\beta}'
=
\boldsymbol{\mathcal{A}}_{\rm LOS}
\boldsymbol{\theta}
-
\boldsymbol{\nabla}\psi_{\rm eff}
(\boldsymbol{\theta}) ,
\end{equation}
with $\boldsymbol{\beta}'=\boldsymbol{\mathcal{A}}_{\rm od}\boldsymbol{\mathcal{A}}_{\rm ds}^{-1}\boldsymbol{\beta}$, $\psi_{\rm eff}(\boldsymbol{\theta})=\psi(\boldsymbol{\mathcal{A}}_{\rm od}\boldsymbol{\theta})$, and $\boldsymbol{\mathcal{A}}_{\rm LOS}=\boldsymbol{\mathcal{A}}_{\rm od}\boldsymbol{\mathcal{A}}_{\rm ds}^{-1}\boldsymbol{\mathcal{A}}_{\rm os}$. This new lens equation is \emph{minimal} in the sense that it trades three line-of-sight distortion matrices for a single one. Note that eq.~\eqref{eq:minimal_LOS_model} is formally equivalent to a main lens with potential $\psi_{\rm eff}$ plus perturbations $\boldsymbol{\mathcal{A}}_{\rm LOS}$ all located in the main lens plane.

In the weak-lensing regime, the line-of-sight distortion matrices are close to unity, the expression of $\boldsymbol{\mathcal{A}}_{\rm LOS}$ can be linearised. Its trace-free part then reads $\gamma_{\rm LOS} = \gamma_{\rm od} + \gamma_{\rm os} - \gamma_{\rm ds}$. This \emph{line-of-sight  (LOS) shear} has a very specific role in the minimal lens equation~\eqref{eq:minimal_LOS_model}, and is therefore expected to be measurable independently of the properties of the main lens. This would make Einstein rings standard shapes: direct probes of $\gamma_{\rm LOS}$.

\section{Measurability of the line-of-sight shear: a proof of concept}

In order to assess the measurability of the LOS shear $\gamma_{\rm LOS}$ on strong-lensing images, we have produced 64 realistic mock images with known line-of-sight perturbations; we then analysed them using the minimal model of eq.~\eqref{eq:minimal_LOS_model} and compared the resulting measurements of $\gamma_{\rm LOS}$ with the input value, namely $\gamma_{\rm od} + \gamma_{\rm os} - \gamma_{\rm ds}$.

\begin{figure}[t]
\centering
\includegraphics[width=\columnwidth]{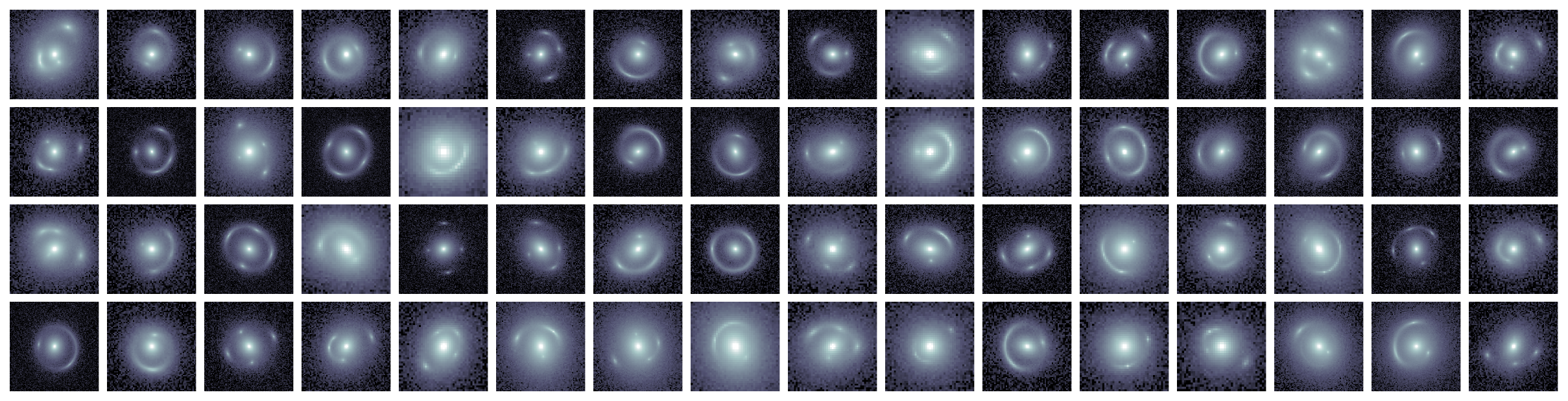}
\includegraphics[width=\columnwidth]{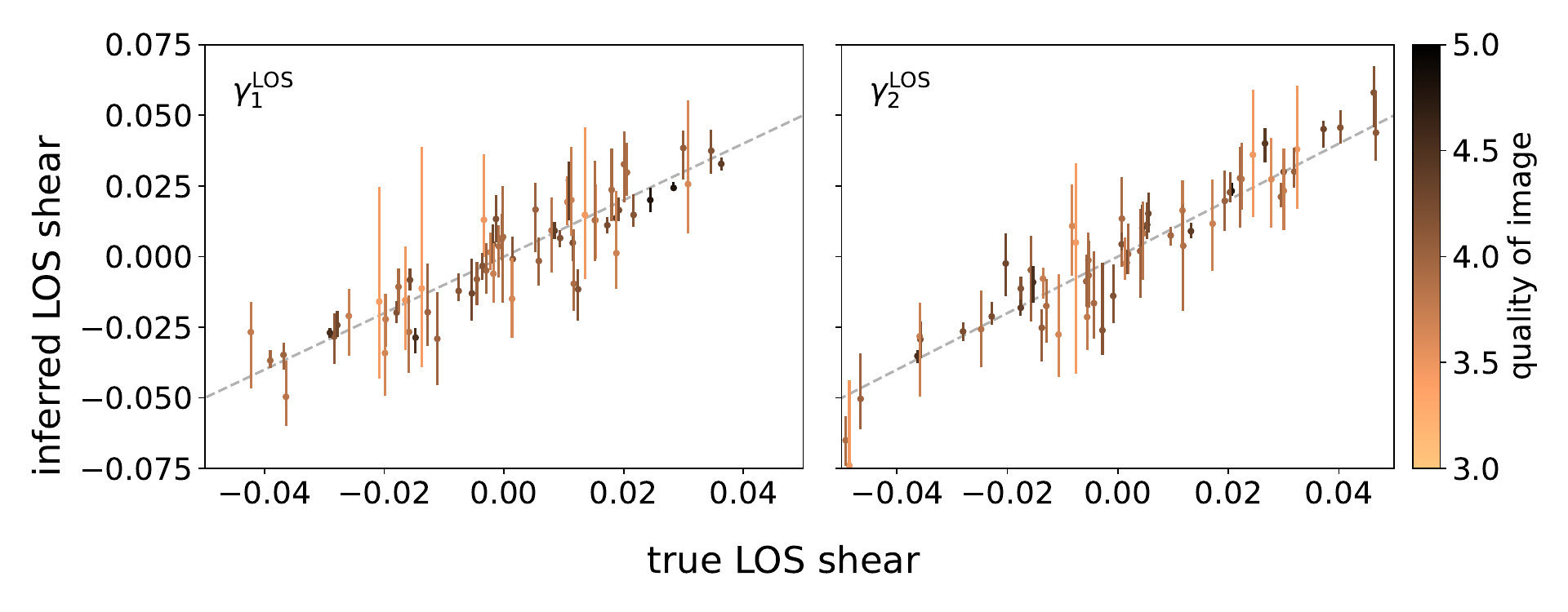}
\vspace*{-1cm}
\caption{\textit{Top panel}: Mock sample of 64 realistic images. \textit{Bottom panel}: Measured components of the LOS shear~$\gamma_{\rm LOS}$ as a function of the input value. Darker colours indicate images with a higher signal-to-noise ratio.}
\label{fig:results_proof_of_concept}
\end{figure}

The mock images were produced using the \href{https://lenstronomy.readthedocs.io/en/latest/}{\textsc{lenstronomy} software},\cite{lenstronomy} with specifications mimicking the image quality of the Hubble Space Telescope. The main lens consists of an elliptical Sérsic profile which represents its baryonic content, superimposed with an elliptical Navarro--Frenk--White profile representing the dark-matter halo of the lens; the centres of both components do not generally coincide. The source is the superposition of several Sérsic profiles. All parameters (masses, ellipticities, scale radii, shears, \ldots) are randomly drawn for each image. The diversity of the resulting mock sample can be observed in the top panel of fig.~\ref{fig:results_proof_of_concept}. 

The results of the mock analysis are reported in the bottom of fig.~\ref{fig:results_proof_of_concept}.\cite{Hogg+2023} We can see that the LOS shear can be accurately measured on those images. The average uncertainty is $\langle\sigma_{\gamma_{\rm LOS}}\rangle=0.01$ on this sample. In other words, the LOS shear can be measured on individual Einstein rings with an average signal-to-noise ratio comparable to unity, which is two orders of magnitude greater than for individual galaxies.

\section{Towards applications in cosmology}

Current photometric galaxy surveys rely on two observables: the fluctuations of the galaxy number density~$\delta_{\rm g}$ on the one hand, and their apparent shapes $\vep$ on the other hand. From those two observables, one can schematically form three two-point correlation functions: $\langle\delta_{\rm g}\times\delta_{\rm g}\rangle$ (galaxy clustering); $\langle\vep\times\vep\rangle$ (cosmic shear); and $\langle\delta_{\rm g}\times\vep\rangle$ (galaxy--galaxy lensing). Cosmological constraints are then extracted from the analysis of this so-called $3\times2\text{pt}$ correlation scheme.

The \emph{Euclid} survey is expected to detect $\mathcal{O}(10^5)$ strong lenses in addition to the $\mathcal{O}(10^9)$ galaxy sample.\cite{Collett_2015} This will add a third observable, the LOS shear~$\gamma_{\rm LOS}$, to $\delta_{\rm g}$ and $\vep$. From this third observable, another three two-point correlation functions can be formed -- $\langle\gamma_{\rm LOS}\times\gamma_{\rm LOS}\rangle$, $\langle\gamma_{\rm LOS}\times\delta_{\rm g}\rangle$ and $\langle\gamma_{\rm LOS}\times\vep\rangle$ -- thereby leading to a $6\times2\text{pt}$ correlation scheme. The additional source of information coming from $\gamma_{\rm LOS}$ is expected to lead to tighter constraints on the cosmological parameters, while mitigating systematic biases due to intrinsic alignments.

Despite its increased signal-to-noise ratio on individual sources, the LOS shear will suffer from its greater sparsity on the sky. Consider the autocorrelation function $\xi^+_{\gamma_{\rm LOS}}\sim \langle\gamma_{\rm LOS}\times\gamma_{\rm LOS}\rangle$ as an example. The estimator of this autocorrelation has three sources of uncertainty:
\begin{equation}
\left\langle (\hat{\xi}^+_{\gamma_{\rm LOS}})^2 \right\rangle
- \left\langle \hat{\xi}^+_{\gamma_{\rm LOS}}\right\rangle^2
= \sigma^2_{\rm cosmic} + \sigma^2_{\rm noise} + \sigma^2_{\rm sparsity} .
\end{equation}
The first term is cosmic variance, due to the fact that we only observe a single Universe; $\sigma_{\rm noise}$ is due to uncertainty on individual measurements of $\gamma_{\rm LOS}$; and $\sigma_{\rm sparsity}$ comes from the fact that we observe a finite number of randomly distributed lenses.

\begin{figure}[t]
\centering
\includegraphics[width=\columnwidth]{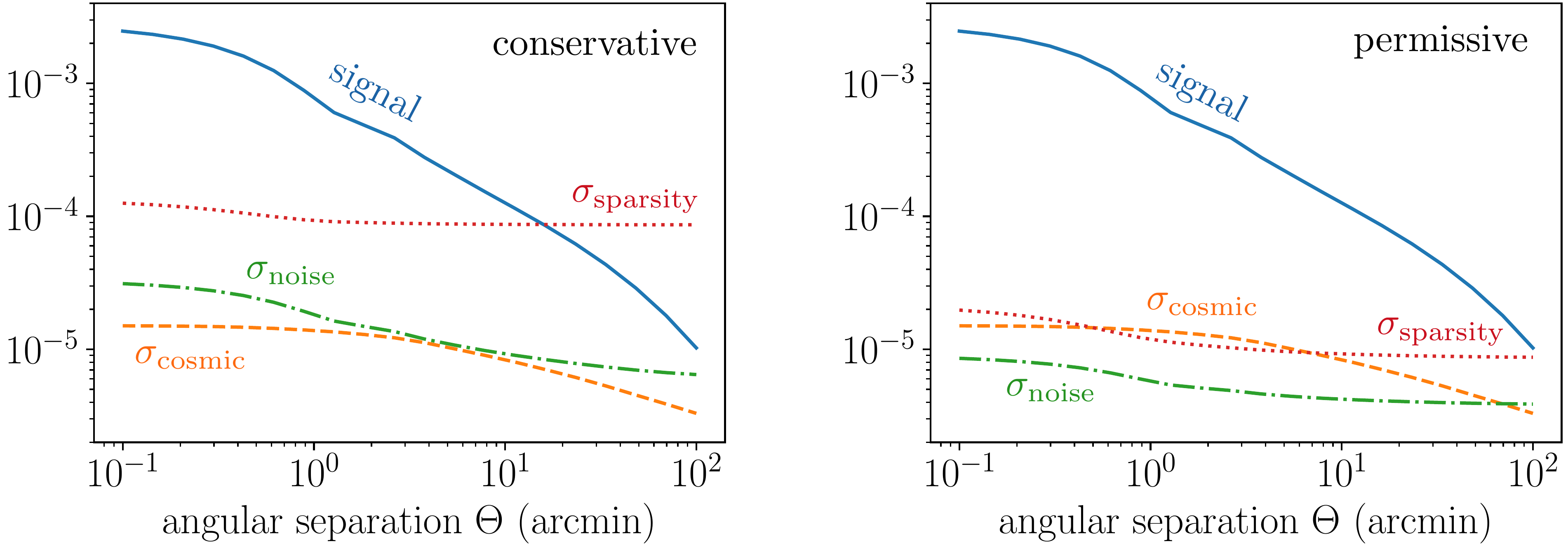}
\caption{Comparison of the expected autocorrelation of the LOS shear~$\xi^+_{\gamma_{\rm LOS}}$ (solid blue), as a function of the angular separation~$\Theta$ of the pairs, with the three sources of uncertainty on this measurement: sparsity (red dotted), noise (green dot-dashed) and cosmic (orange dashed). The left panel corresponds to the conservative scenario (fewer lenses, lower noise) while the right panel is the permissive scenario (more lenses, larger noise).}
\label{fig:xi_+_LOS}
\end{figure}

We compared the expected autocorrelation signal with its uncertainty in two different cases. In the \emph{conservative scenario}, we assumed that about $10\%$ of the expected \textit{Euclid} sample of strong lenses, i.e. about $10^4$ lenses, will be exploitable for $\gamma_{\rm LOS}$ measurements, with an average uncertainty  $\sigma_{\gamma_{\rm LOS}}=0.02$ (twice larger than our proof of concept). In the \emph{permissive scenario}, we assumed that $10^5$ lenses can be exploited, but with a larger uncertainty, $\sigma_{\gamma_{\rm LOS}}=0.05$. Preliminary results are shown in fig.~\ref{fig:xi_+_LOS}. We can see that sparsity variance dominates the uncertainty budget in the conservative scenario, while the permissive scenario is cosmic-variance limited.

\section{Conclusion}

We have demonstrated on realistic mock images that Einstein rings can be used as standard shapes. Specifically, the line-of-sight shear $\gamma_{\rm LOS} = \gamma_{\rm od} + \gamma_{\rm os} - \gamma_{\rm ds}$, due to perturbations besides the main lens, can be measured independently on the properties of the main lens, with a signal-to-noise ratio of order unity. With the expected number of strong lenses observed by the \textit{Euclid} survey, such an observable would lead to a cosmic-variance-limited measurement of cosmic shear, which in addition would be insensitive to intrinsic alignments. The line-of-sight shear of Einstein rings is thus an extremely promising cosmological probe.

\section*{Acknowledgments}

I would like to thank the organisers of the 2024 58\textsuperscript{th} \emph{Rencontres de Moriond}, but also its participants, for making this meeting so wonderful. I acknowledge support from the French \emph{Agence Nationale de la Recherche} through the ELROND project (ANR-23-CE31-0002).

\end{document}